\documentclass[aps,prd,twocolumn,floatfix,superscriptaddress,nofootinbib]{revtex4-1}

\newcommand\mdp{m_{A^\prime}}
\newcommand\Adp{A^\prime}
\newcommand\rtil{\tilde{r}}
\usepackage{amsmath}
\usepackage{graphicx,bm}
\usepackage[utf8]{inputenc} 
\usepackage[colorlinks,citecolor=blue]{hyperref}
\begin{document}

\title{New constraint from supernova explosions on light particles beyond the Standard Model}

\author{Allan Sung}
\email{allan93161@gmail.com}
\affiliation{Institute of Physics, Academia Sinica, 
  Taipei, 11529, Taiwan}
\affiliation{Department of Physics, National Taiwan University,
  Taipei, 10617, Taiwan}

\author{Huitzu Tu}
\email{huitzu2@gate.sinica.edu.tw}
\affiliation{Institute of Physics, Academia Sinica, 
  Taipei, 11529, Taiwan}

\author{Meng-Ru Wu}
\email{mwu@gate.sinica.edu.tw}
\affiliation{Institute of Physics, Academia Sinica, 
  Taipei, 11529, Taiwan}
\affiliation{Institute of Astronomy and Astrophysics, Academia Sinica, 
  Taipei, 10617, Taiwan}

\begin{abstract}
We propose a new constraint on light (sub-GeV)
particles beyond the Standard Model
that can be produced inside the 
proto-neutron star core resulting from
the core-collapse supernova explosion.
%that couples to the SM sector
%electromagnetically.  
It is derived by demanding that the 
energy carried by exotic particles 
being transferred to the progenitor
stellar envelopes must not exceed the
explosion energy of $\lesssim 2\cdot 10^{51}$~erg
of observed supernovae.
We show specifically that for the case of a  
dark photon which kinetically mixes with the SM photon and decays predominantly to an $e^\pm$ pair, a
smaller mixing parameter of one order of magnitude
below the well-established supernova cooling bound can be excluded.
Furthermore, our bound fills the gap between the 
cooling bound and the region constrained by
(non)observation of $\gamma$ rays produced from 
supernovae for dark photons lighter than $\sim 20$~MeV.
Our result also rules out the possibility of
aiding successful supernova explosions by transferring energy from the supernova core to 
the shock with exotic particles.
\end{abstract}

\date{\today}
\maketitle

%%%%%%%%%%%%%%%%% BODY OF PAPER %%%%%%%%%%%%%%%%%%

\section{Introduction}\label{sec-intro}
The Standard Model (SM) of particle physics has been
the most successful theory that describes the
fundamental properties and interactions between
elementary particles. 
However, various hints from either the theoretical
considerations or the cosmological and
astrophysical observations point to the
possibility that it is not a complete
theory and new particles beyond the SM (bSM) that
only couple to the SM sector very weakly may exist.

Among the imperative searches and constraints of bSM
particles, one important criterion comes from
the observation of electron antineutrinos
($\bar\nu_e$) associated 
with the seminal core-collapse supernova (CCSN) event,
SN1987A. 
The observed $\bar\nu_e$ burst duration of about 
12~s, with individual energies up to $40~{\rm MeV}$,
as well as the integrated total energy 
$\sim 5\cdot 10^{52}$~erg~\cite{Hirata:1987hu,Bionta:1987qt,Alekseev:1988gp,Sato:1987rd,Spergel:1987ch,Bahcall:1987,Burrows:1987,Loredo:2001rx},
strongly supported the standard picture of neutrino
cooling of the proto-neutron star (PNS): The 
total gravitational binding energy, 
$E_G\sim 3\cdot 10^{53}$~erg, 
released while forming a compact PNS with a mass 
$M_{\rm PNS}\sim 1.4$~$M_\odot$ and radius 
$R_{\rm PNS}\sim 10$~km is roughly equipartitioned by 
all six flavors of (anti)neutrinos.
Consequently, any bSM particles that can be produced
inside the PNS and escape by taking away an energy
comparable to $E_G$
would have shortened the observed timescale of the 
$\bar\nu_e$ burst to be incompatible with the
observation~\cite{Raffelt:1990yz}.

Constraints on various light bSM particles that may be 
produced in the hot and dense PNS core, based on the above
argument, have been
considered exhaustively in the literature, notably
the axions~\cite{Raffelt:1987yt,Turner:1987by,Mayle:1987as,Brinkmann:1988vi,Janka:1995ir}, right-handed neutrinos~\cite{Raffelt:1987yt,Raffelt:2011nc,Arguelles:2016uwb}, Majorons~\cite{Farzan:2002wx},
Kaluza-Klein gravitons~\cite{Hanhart:2000er,Hanhart:2001fx,Hannestad:2003yd}, Kaluza-Klein dilatons~\cite{Hanhart:2000er}, unparticles~\cite{Hannestad:2007ys,Freitas:2007ip}, 
dark photons~\cite{Dent:2012mx,Rrapaj:2015wgs,Chang:2016ntp,Hardy:2016kme}, dark matter~\cite{Chang:2018rso,Guha:2015kka,Guha:2018mli}, dilaton~\cite{Ishizuka:1989ts}, saxion~\cite{Arndt:2002yg}, Goldstone bosons~\cite{Keung:2013mfa,Tu:2017dhl}, etc.
Ideally, one should perform numerical simulations as in Refs.~\cite{Keil:1996ju,Hanhart:2001fx,Fischer:2016cyd} to study the effects of a light bSM particle on the neutrino burst signal. 

Other than affecting the PNS cooling, 
bSM particles produced inside the PNS may directly decay to 
photons, or indirectly produce the 511~keV lines via
the pair-annihilation by first decaying into $e^\pm$, outside the 
surface of the progenitor stars, $R_\ast\simeq 10^{14}$~cm. 
The (non)observation of $\gamma$ rays associated with SN1987A,
as well as the observed flux of 511~keV photons from the 
Milky Way has been used to put constraints on bSM particles
that couple electromagnetically to the 
SM sector~\cite{Kazanas:2014mca,Jaeckel:2017tud,DeRocco:2019njg}. 
Such derived bounds mostly complement those from
the PNS cooling, because for bSM particles to decay
outside $R_\ast$, the required coupling to the SM sector is usually
not large enough to affect the PNS cooling.

In this paper, we propose a new constraint
that bridges those from the PNS cooling and the
$\gamma$-ray (non)observation. 
Our new constraint is based on a very basic fact:
The known explosion energy of the CCSN
of a progenitor star with $10$~$M_\odot\lesssim M_\ast \lesssim 20$~$M_\odot$
is $\simeq 1$~B, where B stands for bethe $\equiv 10^{51}$~ergs~\cite{Nomoto:2013oal,Bruenn:2014qea}.
Most of this energy is carried by the
kinetic energy of the expanding ejecta, 
with a mass of $\sim \mathcal{O}(10)$~$M_\odot$
and a velocity of $\sim 0.01\, c$, when we observed
the emitted (quasi-)thermal photons at $\geq\mathcal{O}(1)$~d after the core bounce~\cite{Falk:1977xx}.
In the absence of bSM physics, the prevalent theory
is that the neutrinos emitted from the PNS within
$\sim 1$~s after the core bounce, can
deposit a few percent of their energy to the stalled
shockwave at $\sim \mathcal{O}(10^2)$~km to 
revive it~\cite{Janka:2012wk}.
The shock then wipes out the outer stellar envelopes 
at a speed of $\leq 0.1\, c$, giving rise to the 
observed explosion.

\begin{figure}[htbp]
  \centering \includegraphics[width=0.8\linewidth]{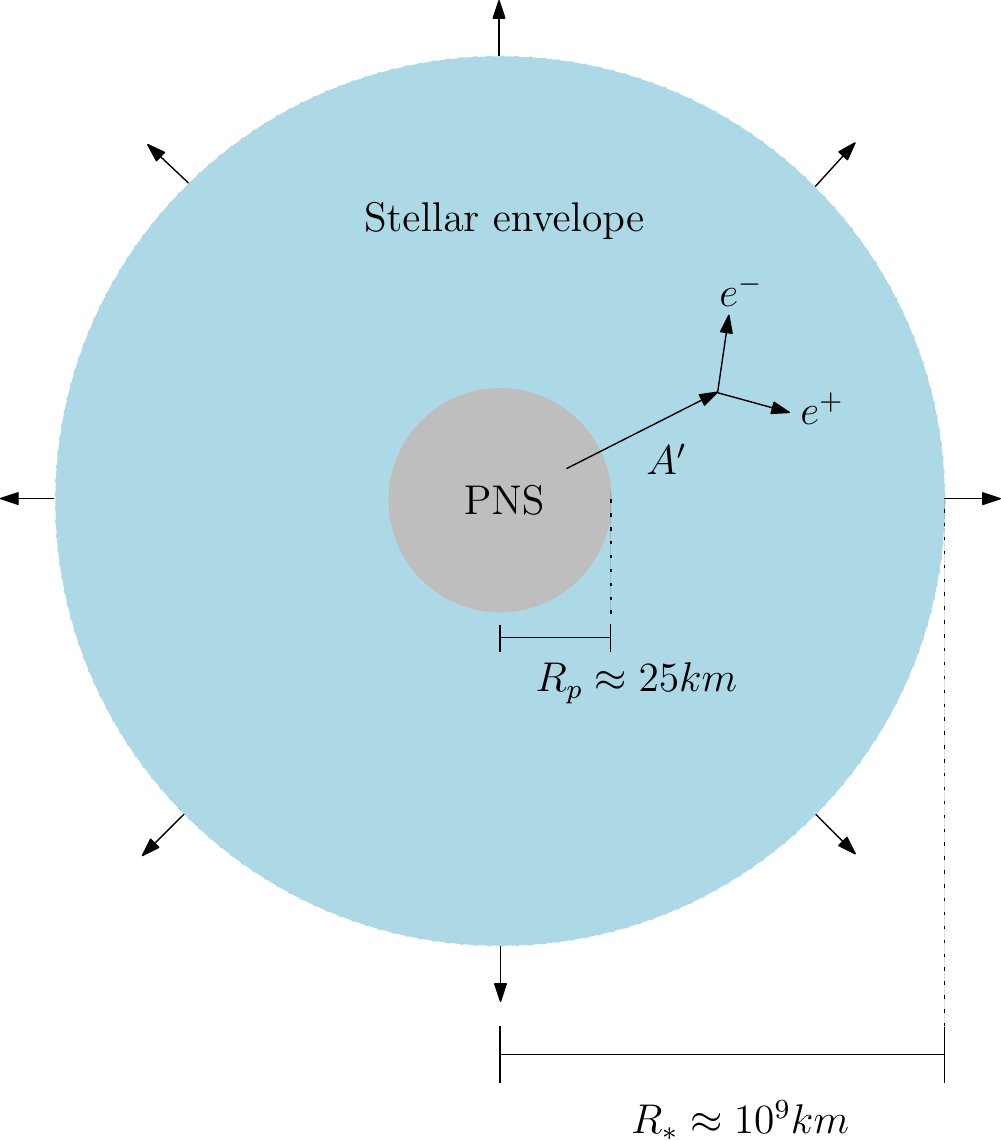}
  \caption{A schematic plot showing the energy deposition of the bSM particles produced from
  the PNS within a radius $R_p$ into the
  stellar layers of the progenitor star with
  a radius $R_\ast$.
  Here we illustrate it with the example of 
  a dark photon ($\Adp$) decaying into an 
  $e^\pm$ pair.
  \label{fig:sche}}
\end{figure}

However, if bSM particles produced from the PNS
can transfer the energy that they carry 
into the stellar envelopes or the shocked material
before leaving the progenitor star, 
they would serve as a new energy
source contributing to the total explosion energy
(see Fig.~\ref{fig:sche} for a schematic plot).
As a result, if this energy deposition mediated by
bSM particles exceeds the observed explosion energy,
after subtracting the gravitational binding of the
stellar envelopes, such a bSM particle is then ruled
out by CCSN observation. 

Before working out a specific example, we first 
demonstrate analytically how these new bounds can 
improve the constraint derived from the PNS cooling.
A well-known analytic criterion formulated by
G. Raffelt of such states the following:
For a novel cooling agent $X$ that free-streams after
production, its specific energy loss
$\dot\varepsilon$ is bounded by~\cite{Raffelt:1990yz}
\begin{equation}\label{eq:Raffelt}
  \dot{\varepsilon}_X \lesssim \frac{L_\nu}{M_{\rm PNS}} \simeq 10^{19}~{\rm erg}\, {\rm g}^{-1}\, {s}^{-1}\, ,    
\end{equation}
with $L_\nu \sim E_G/10 \simeq 3 \cdot 10^{52}$~erg~s$^{-1}$ 
being the energy luminosity of all (anti)neutrinos and
$\dot\varepsilon_X$ being evaluated at a typical core
condition at $\sim 1$~s after the core bounce, 
with a temperature of $\simeq 30~{\rm MeV}$
and a density of 
$\simeq 3 \cdot 10^{14}~{\rm g}\, {\rm cm}^{-3}$. 

The upper bound of the observed explosion energy of CCSNe
associated with progenitor stars with
zero-age main-sequence (ZAMS) masses between $10$ and
$20$~$M_\odot$ is mostly under $E_{\rm expl}=2$~B (see e.g., the compilations in 
Refs.\citep{Nomoto:2013oal,Bruenn:2014qea,Ebinger:2018fkw}),
while the typical binding energy of the stellar
envelopes is $E_b\lesssim 1$~B (see later in this paper for details). 
Therefore, our proposed new constraint can be expressed by
\begin{equation}\label{eq:ana-new}
  K \cdot \dot\varepsilon_X \lesssim \frac{E_{\rm expl}+E_b}{\Delta t \cdot M_{\rm PNS}} \lesssim 10^{17}~{\rm erg}\, {\rm g}^{-1}\, {s}^{-1}\, ,
\end{equation}
where $\Delta t\simeq 10$~s,
and $0< K \leq 1$ denotes the efficiency of energy
transfer into the region between a radius
$R_p$, within which the particle $X$ can be produced efficiently, 
and $R_\ast$.
Comparing Eqs.~\eqref{eq:Raffelt} and \eqref{eq:ana-new}, it is obvious that
the new bound can exclude the bSM particle 
whose emissivity is $\sim$ 2 orders of magnitude
less than the one constrained by the PNS cooling,
for cases where $K\sim 1$.
For the rest of the paper, we consider a specific
example of the dark photon that decays predominantly to
an $e^\pm$ pair. 

\section{New constraint on dark photon}\label{sec-cons}

We consider the minimal extension of the SM with a $U (1)^\prime$ dark sector. 
The dark photon ($A^\prime$) is the gauge boson of the broken $U (1)^\prime$ symmetry which kinetically mixes with the hypercharge boson.
When the dark photon mass is much smaller than the electroweak symmetry breaking scale, the mixing is effectively only with the photon ($A$).
The effective Lagrangian for the photon--dark photon system is
(see, e.g., Ref.~\cite{Feng:2016ijc} for the transformation from the dark photon gauge eigenstates to the mass eigenstates)
\begin{eqnarray}
    \mathcal{L} &=& - \frac{1}{4} F_{\mu \nu} F^{\mu \nu} -  \frac{1}{4} F^\prime_{\mu \nu} F^{\prime \mu \nu} + \frac{1}{2} \mdp^2 \Adp_\mu A^{\prime \mu} \nonumber \\
    && - e\, \sum_f q_f (A_\mu + \epsilon \Adp_\mu)\, \bar{f} \gamma^\mu f\, .   
\end{eqnarray}
Here $f$ is a SM fermion with electric charge $q_f$, and $\mdp$ and $\epsilon$ are the mass and the kinetic mixing parameter of the dark photon in the physical basis, respectively.
Strategies for dark photon searches at colliders and fixed-target experiments, existing constraints on $(\epsilon, \mdp)$, as well as anticipated sensitivities of planned experiments, can be found in the reports~\cite{Alexander:2016aln,Beacham:2019nyx}.
 
The in-medium physical eigenstates are quite distinct from those in vacuum due to the presence of the photon polarisation tensor $\Pi = \Pi_R + i \Pi_I$ (see, e.g., Ref.~\cite{Braaten:1993jw}) in the inverse propagator matrix of the photon-dark photon system.
As a consequence, in hot or dense stars the collective effects of the stellar plasma can significantly change the dark photon production rate~\cite{An:2013yfc,Redondo:2013lna}.
References~\cite{Chang:2016ntp,Hardy:2016kme} found that plasma effects in the PNS qualitatively weaken the supernova cooling bound at dark photon masses below $\sim 10~{\rm MeV}$.

In this work, we calculate the dark photon production rate, following closely Refs.~\cite{Chang:2016ntp,Hardy:2016kme}.
For a dark photon weakly coupled to the thermal bath---i.e., when $\epsilon \ll 1$---one can invoke the in-medium effective kinetic mixing parameter
\begin{equation}
   \epsilon^2_m = \frac{\epsilon^2}{(1 - \Pi_R / \mdp^2)^2 + (\Pi_I / \mdp^2)^2}\, ,   
\end{equation}
for the transverse ($T$) and the longitudinal ($L$) polarisations separately.
In CCSNe, the real part of the photon polarisation tensor, $\Pi_{R \vert L, T}$, is dominantly generated by the electrons, which are relativistic and degenerate inside the neutrino sphere $R_\nu$.
The imaginary part $\Pi_{I \vert L, T}$ is determined mainly by the rates of the nuclear bremsstrahlung and the Compton scattering processes.
Transversely and longitudinally polarised dark photons can thus be produced in the corresponding channels ($p n \rightarrow p n \Adp$, $p p \rightarrow p p \Adp$, and $\gamma e^- \rightarrow e^- \Adp$) through the effective in-medium mixing with the photon. Since for $\epsilon_m$, the condition $\Pi_I \ll \Pi_R$ generally holds throughout the PNS environment, the resonant emission of longitudinal dark photons is open for $\mdp < \omega_p$, where $\omega_p$ is the photon plasma mass.  
Resonant emission of transverse dark photons, on the other hand, is only possible for $\mdp$ in a narrow range around $\omega_p$.   
 
Dark photons are reabsorbed in the supernovae mainly by the decay process $\Adp \rightarrow e^+ e^-$ when it is kinematically allowed. 
As pointed out in Ref.~\cite{Hardy:2016kme},
in the PNS core region, dark photon decay is prevented due to the high electron chemical potential, unless $\mdp$ is larger than twice the effective electron mass in the plasma~\cite{Braaten:1991hg}.
In this work, we are interested in the case in which the dark photon can escape the production region and decay freely in the stellar layers.
The produced $e^\pm$ then quickly interact with the medium and
lose their kinetic energy of $\sim 10$--$100$~MeV to the
surroundings in a length scale much shorter than $R_\ast$~\cite{Gould:1972xx}. 
This effectively leads to an efficient transfer of the thermal
energy from the PNS core region to the 
stellar envelope [$K \simeq 1$ in Eq.~(\ref{eq:ana-new})] . 

\begin{figure}[tbp!]
  \centering \includegraphics[width=\linewidth]{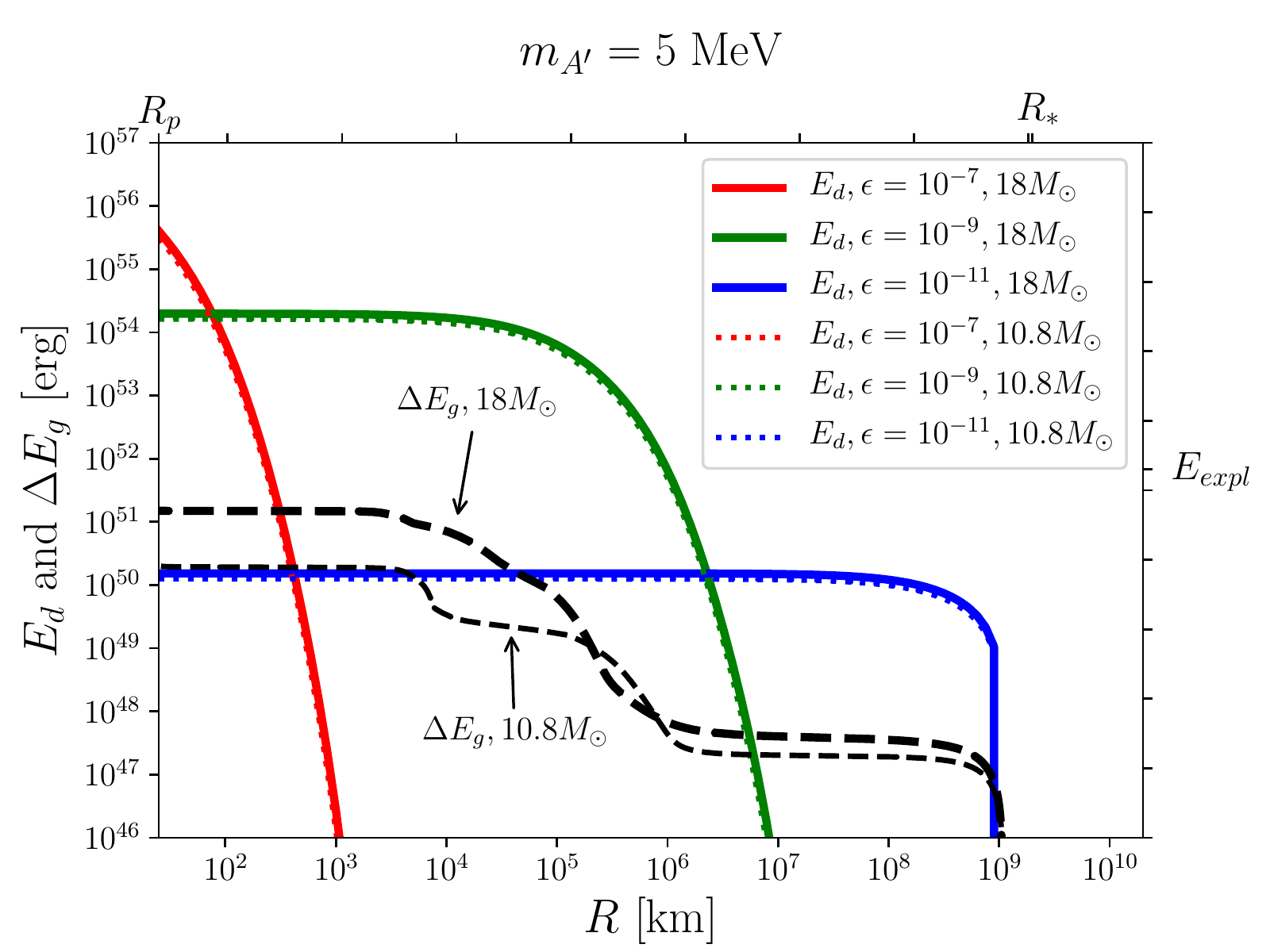}
  \caption{
  Energy deposition $E_d (R)$ by dark photons to stellar envelopes outside radius
  $R$, for various dark photon parameters and for supernovae with progenitor masses of $18\, M_\odot$ (thick solid curves) and
  $10.8\, M_\odot$ (thin dotted curves). 
  Also shown are the corresponding gravitational binding energy $\Delta E_g (R)$ outside $R$ in both cases (thick and thin dashed curves).
  \label{fig:Ecomp}}
\end{figure} 
 
For a given dark photon mass $\mdp$ and kinetic coupling $\epsilon$, the total energy carried by the dark photons to a distance $R \geq R_p$ is calculated by~\cite{Chang:2016ntp}
\begin{eqnarray}
   L_{\Adp} (R, \mdp, \epsilon) &=& \sum_{L, T}\, \int_{r=0}^{R_p} \int^\infty_{\omega=\mdp} d r\, d \omega\, 4 \pi r^2\, e^{- \tau_{L, T} (r, \omega, R)} \nonumber \\ 
   && \hspace{-2.7cm} \cdot\, \frac{\omega^3 v}{2 \pi^2}\, e^{- \frac{\omega}{T (r)}}\, \epsilon^2_{m \vert L, T} (r, \omega)\, \cdot \left[\Gamma_{iBr \vert L, T} (r, \omega) + \Gamma_{sC \vert L, T} (r, \omega) \right]\, , \nonumber \\
   &&
\end{eqnarray} 
assuming that the nucleons and electrons are in local thermal equilibrium at temperature $T(r)$.
Under this condition the total production and the total absorption rate of SM photons of energy $\omega$ are related by $\Gamma_{\rm prod} = e^{ - \omega / T (r)}\, \Gamma_{\rm abs}$, where $\Gamma_{\rm abs}$ is determined by $\Gamma_{iBr}$ and $\Gamma_{sC}$, the inverse bremsstrahlung and the semi-Compton process rates, respectively.
For $\Gamma_{iBr}$, we adopt the soft-radiation approximation and neglect many-body effects in the nuclear medium, as Ref.~\cite{Rrapaj:2015wgs}.
Therefore, dark photons are created through in-medium kinetic mixing with the SM photons at the rate $\Gamma^\prime_{{\rm prod} \vert L, T} = e^{- \omega / T (r)}\, \epsilon^2_{m \vert L, T}\, \Gamma^{(\rm in\, eq.)}_{{\rm abs} \vert L, T}$.
The photon velocity in medium is $v = \sqrt{1 - \mdp^2 / \omega^2}$.
The term $e^{- \tau (r, \omega ,R)}$ takes into account dark photon attenuation between radius $r$ and $R$.
We calculate the optical depth for a dark photon produced at radius $r$ with energy $\omega$, which travels radially outward to $R$ by
\begin{equation}
  \tau_{\rm radial\, out} (r, \omega, R) = \left[\int^{R_p}_r \frac{d \rtil}{v}\, \Gamma^\prime_{\rm abs \vert L, T} + \frac{R - R_p}{v} \cdot \Gamma^\prime_{e^+ e^-} \right]\, , 
\end{equation}
and include a correction factor to relate $\tau (r)$ to $\tau_{\rm radial\, out} (r)$ as suggested by Ref.~\cite{Chang:2016ntp}.
The dark photon absorption rate $\Gamma^\prime_{\rm abs}$ receives contributions from the inverse bremsstrahlung processes, semi-Compton scattering, and decay to $e^\pm$ pairs.
We have checked that outside $R_p$, Pauli blocking can be ignored, and one can use the decay rate in vacuum for $\Gamma^\prime_{e^+ e^-}$.

The supernova cooling bound is determined by $L_{\Adp} (R_p) \leq  L_\nu$ [cf. Eq.~(\ref{eq:Raffelt})] in the dark photon $(\mdp, \epsilon)$ parameter space.
Our new bound, Eq.~(\ref{eq:ana-new}), is by requiring that 
the energy deposited by the decay of $\Adp$ between $R_p$ and $R_\ast$ be
smaller than the sum of the observed SN explosion
energy and the total gravitational binding energy
between these two radii: 
\begin{equation}\label{eq:ediff}
  E_d(R_p)\equiv \left[ L_{\Adp} (R_p) - L_{\Adp} (R_\ast) \right] \cdot \Delta t \leq E_{\rm expl} + \Delta E_g (R_p)\, .
\end{equation}
Here $\Delta E_g (R) \equiv E_g (R_\ast) - E_g (R)$, with 
\begin{equation}
    E_g (R) \equiv \int^R_0 d r\, \frac{G \rho (r)\, M_{\rm enc} (r)}{r}\, 4 \pi r^2\, ,
\end{equation}
the gravitational binding energy inside radius $R$, 
where $M_{\rm enc} (r)$ is the total mass enclosed in the region inside $r$. 
We fix the emission duration $\Delta t = 10~{\rm s}$, which is the typical timescale of the PNS cooling.\footnote{Since the quantity $E_{\rm expl}+\Delta E_g(R_p)$ on the rhs of Eq.~\eqref{eq:ediff} is only $\sim 10^{51}$~erg (see below),
much smaller than the total binding energy of the PNS, $E_G\sim 3\cdot 10^{53}$~erg, 
for dark photons that just carry and deposit 
an energy $E_d(R_p)\gtrsim E_{\rm expl}+\Delta E_g(R_p)$, they would only alter the cooling behavior of the PNS core by $\sim 1\%$.
Therefore, the PNS cooling timescale of $\sim 10$~s should not be affected, and our derived bound based on $\Delta t=10$~s is robust.}
Note that in Eq.~\eqref{eq:ediff}, we have neglected the kinetic energy of the shocked material, as well as
that of the stellar envelope, which contribute at most $\sim 10\%$ of
$E_{\rm expl}$.

The dark photon deposited energy $E_d(R_p)$ and
the gravitational binding energy of the stellar envelope $\Delta E_g(R_p)$ depend on the 
structure of the PNS and the mass of the stellar progenitor.
We examine two cases using the radial profile
of the mass density, temperature, electron fraction, and electron chemical potential
obtained by SN simulations of progenitor stars
with $10.8\, M_\odot$ and $18\, M_\odot$ masses~\cite{Fischer:2009af}, chosen at $t = 1~{\rm s}$ after the core bounce.
As those SN simulations do not contain the structure of the outermost hydrogen layer of the progenitor star, we
extend the profile to $R_\ast$ using the pre-SN
structure provided by Ref.~\cite{Muller:2016ujh}.
For both cases, we have used the same $R_p=25$~km (slightly larger than $R_\nu$) 
so as to encompass all the dark photon resonant production sites.

Figure~\ref{fig:Ecomp} shows the comparison of 
$E_d(R)$ calculated with $\mdp=5$~MeV and a few
selected $\epsilon=10^{-7}$, $10^{-9}$, and $10^{-11}$,
to $\Delta E_g(R)$ for both progenitor masses.
Different progenitor masses only lead to distinct $\Delta E_g(R)$ for $R>R_p$,
but not $E_d(R)$, because the PNS structure is almost independent of the progenitor mass. 
For a given $\mdp$, dark photons with larger (smaller) $\epsilon$ 
carry more (less) energy away from the PNS and
decay to $e^\pm$ at smaller (larger) radii above $R_p$.
For $\epsilon=10^{-7}$ and $10^{-9}$, the energy
deposition by the dark photon decay far exceeds the
gravitational binding energy of the envelope by
several orders of magnitude and can therefore be
ruled out by our criterion.
With $\epsilon=10^{-11}$, dark photons only
carry $\sim 10^{50}$~erg of energy away from the PNS
and therefore cannot be ruled out by our constraint.

\begin{figure}[tbp!]
  \centering \includegraphics[width=\linewidth]{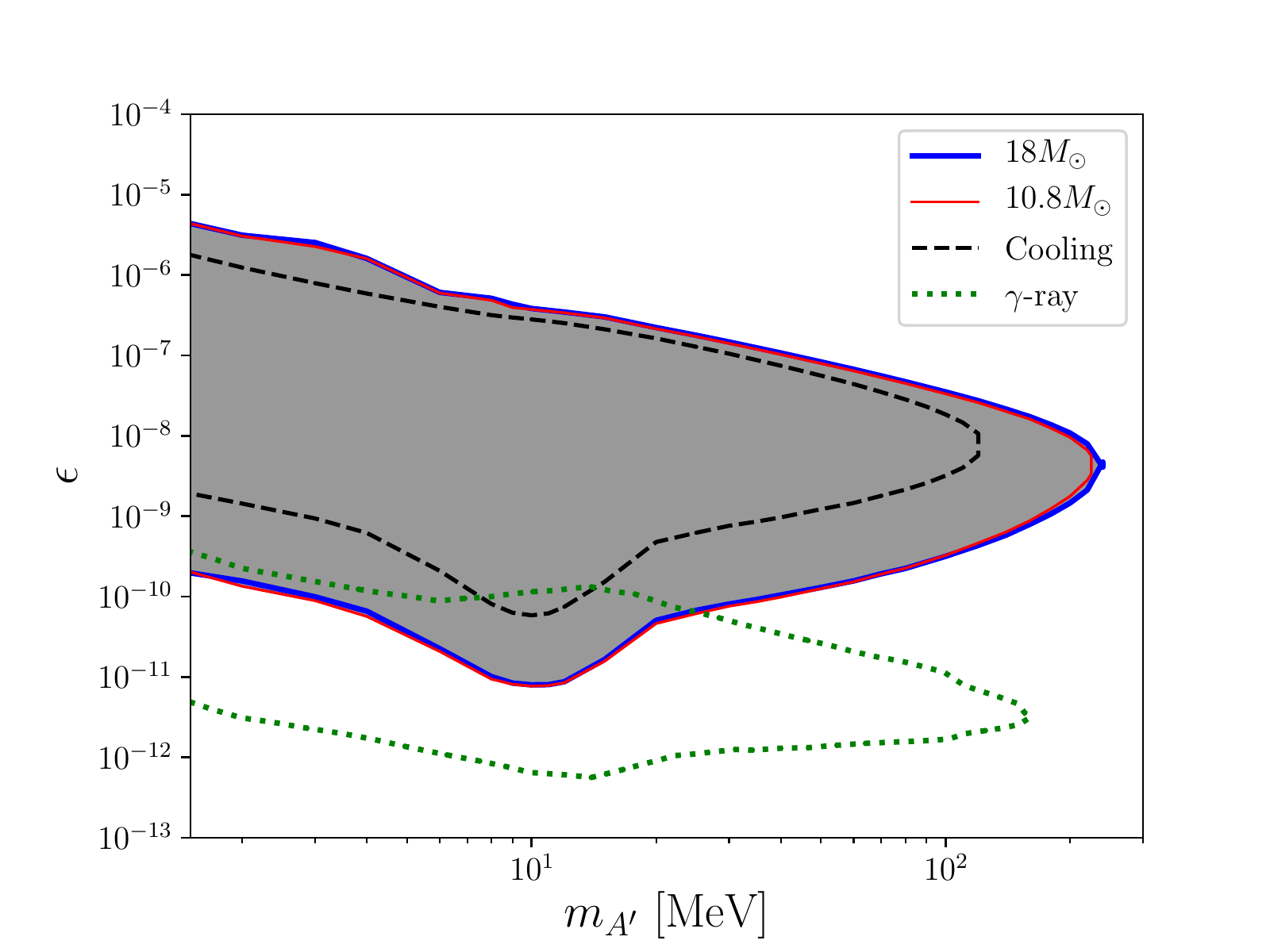}
  \caption{Shaded region: excluded parameter space of dark photon derived using the observed SN explosion energy for progenitor
  masses of $18$ and $10.8\, M_\odot$.
  The black dashed curve shows the bound determined by the PNS cooling argument for $18\, M_\odot$.
  Also shown is the excluded region inferred from the (non)observation of $\gamma$ rays (dotted green curve), taken from Ref.~\cite{DeRocco:2019njg}. 
  \label{fig:contour}}
\end{figure}

In Fig.~\ref{fig:contour}, we show the contour
plot for regions excluded by our new constraint, and that excluded by the PNS cooling, computed as
aforementioned. In addition, we show the excluded region by the $\gamma$-ray (non)observation from
Ref.~\cite{DeRocco:2019njg}.
The regions excluded by the observed SN explosion energy are 
nearly identical for both the $10.8\, M_\odot$
and $18\, M_\odot$ progenitors because
$E_d(R_p)$ are almost the same and
$\Delta E_g(R_p)\ll E_d(R_p)$ 
for most of the excluded region other than
those very close to the boundary
(see Fig.~\ref{fig:Ecomp}).
Their shapes closely follow and enclose that from the PNS
cooling constraint. 
Their upper boundaries denote the $\epsilon$ value for which dark photons of mass $\mdp$ are produced copiously but also reabsorbed strongly inside radius $R_p$.   
Besides, as expected by our analytic estimate, 
this new consideration extends the excluded region to 
a lower $\epsilon$ by roughly 1 order of magnitude (which corresponds to a factor of $\sim 100$ in terms of dark photon
emissivity) for a given $\mdp$. Note that as it largely overlaps with the $\gamma$-ray bound in the small-$\epsilon$ 
regime, they form together a robust bound covering nearly 6 orders of magnitude for $\mdp\lesssim 20$~MeV.

\section{Discussions}\label{sec-dis}
We have shown that the observed explosion energy of
the CCSNe can be used to derive important
constraints on light bSM particles that may be
copiously produced from the PNS core. 
For dark photons that kinetically mix with SM photons, we show
that our new bound excludes a larger parameter space 
than that derived 
using the observed neutrino burst from SN1987a.
Moreover, it overlaps with the region 
recently obtained using the
(non)observation of $\gamma$ rays produced by supernovae. 
Therefore, all three constraints together exclude
a large range of parameter space that is 
not accessible by current terrestrial experiments
or by cosmological observation.
Although we have only considered the explicit example of dark photons, constraints on other bSM
particles such that may effectively transfer energy
from the PNS to the stellar layers---e.g., the sub-GeV
axion-like particles~\cite{Jaeckel:2017tud} and
MeV sterile neutrinos~\cite{Rembiasz:2018lok}---can be similarly derived.

Besides, in new physics models, it is sometimes speculated that the light bSM particles escaping the PNS may deposit energy into the gain
region behind the stalled supernova shock to revive it and facilitate supernova explosions, in case neutrino heating is not effective. 
Such a scenario would typically require that light bSM particles provide an additional heating rate of $\sim$ a few times $10^{51}$~erg~s$^{-1}$, similar to 
that from neutrino heating (see, e.g., Fig.~1 in Ref.~\cite{OConnor:2018tuw}).
As this rate is about 1 order of magnitude smaller than the luminosity $L_\nu$ used to derive the SN
cooling bound, the corresponding parameter space cannot be excluded by the cooling bound.
However, our new constraint dictates that the average luminosity of any bSM particles emitted from PNS cannot exceed $\sim 1\%$ of $L_\nu$, if they can deposit energy above PNS 
[see Eqs.~\eqref{eq:Raffelt} and \eqref{eq:ana-new}]. It thus rules out the possibility of a
light bSM particle reviving the SN shock, because otherwise the continuous energy
injection to stellar layers during the PNS cooling will lead to explosions 
that are too energetic.

Several uncertainties may affect the exact 
excluded region derived with the simple argument
presented in this work for dark photons. 
For example, improved description of 
the dark photon emission from the nuclear
bremsstrahlung beyond the soft radiation approximation adopted here, incorporating the somewhat uncertain condition of the PNS core temperature, density, and composition (see, e.g., Ref.~\cite{Mahoney:2017jqk}), as well as the time-dependence of the PNS structure and the
stellar envelope profile, may introduce some minor corrections.
Nevertheless, we would like to emphasize that 
these uncertainties would affect all the derived bounds, including that from the PNS cooling and the $\gamma$-ray (non)observation. Therefore, the main message of this paper remains solid despite these uncertainties: the observed explosion energy of core-collapse supernovae places improved 
constraint on bSM particles that 
are able to transfer energy efficiently from the PNS core to the stellar mantle.

On the other hand, a detailed SN light-curve modelling
taking into account ejecta driven by bSM particles can potentially 
provide even better constraints. 
For instance, even if the bSM particles only unbind the outermost part of the stellar envelope 
with an energy smaller than $E_{\rm expl}$ (see, e.g., the
case with $\epsilon=10^{-11}$ in Fig.~\ref{fig:Ecomp}), the standard neutrino-driven
mechanism can still work to eject the entire inner layers. Depending on their relative velocity,
those two ejecta may collide at times of a few days after the core collapse and lead to very 
luminous events not compatible with observations.
The hydrogen layer of the stellar envelope may
also be driven off by the energy deposition
from bSM particles with a speed much larger than
typical SN ejecta velocity, resulting in an
electromagnetic precursor prior to the main
supernova peak lights, or reducing the 
line feature of hydrogen. 
All these aspects require more dedicated work beyond the 
scope of this paper and deserve further
exploration.

\begin{acknowledgments}
  The authors thank the anonymous referees for providing useful comments and suggestions.
  This work is partly supported under Grant No.~107-2119-M-001-038 from the Ministry of Science and Technology, Taiwan, 
  and from the National Center for Theoretical Sciences.
\end{acknowledgments}

%\bibliography{reference}
%

\end{document}